# Large Exchange Bias Effect and Coverage-Dependent Interfacial Coupling in CrI$_3$/MnBi$_2$Te$_4$ van der Waals Heterostructures

Zhe Ying,[†,#] Bo Chen,[†,#] Chunfeng Li,[†] Boyuan Wei,[†] Zheng Dai,[†] Fengyi Guo,[†] Danfeng Pan,[‡,§] Haijun Zhang,[†] Di Wu,[†] Xuefeng Wang,[‡] Shuai Zhang,[†,∥,*] Fucong Fei,[†,∥,*] and Fengqi Song[†,∥,*]

[†]*National Laboratory of Solid State Microstructures, Collaborative Innovation Center of Advanced Microstructures, and School of Physics, Nanjing University, Nanjing, 210093, China*

[‡]*National Laboratory of Solid State Microstructures, Collaborative Innovation Center of Advanced Microstructures, and School of Electronic Science and Engineering, Nanjing University, Nanjing, 210093, China*

[§]*Microfabrication and Integration Technology Center, Nanjing University, Nanjing, 210093, China*

[∥]*Atom Manufacturing Institute, Nanjing, 211806, China*

[#]These authors contributed equally to this work.
[*]szhang@nju.edu.cn
[*]feifucong@nju.edu.cn
[*]songfengqi@nju.edu.cn

**ABSTRACT:** Igniting interface magnetic ordering of magnetic topological insulators by building a van der Waals heterostructure can help to reveal novel quantum states and design functional devices. Here, we observe an interesting exchange bias effect, indicating successful interfacial magnetic coupling, in $CrI_3$/$MnBi_2Te_4$ ferromagnetic insulator/antiferromagnetic topological insulator (FMI/AFM-TI) heterostructure devices. The devices originally exhibit a negative exchange bias field, which decays with increasing temperature and is unaffected by the back-gate voltage. When we change the device configuration to be half-covered by $CrI_3$, the exchange bias becomes positive with a very large exchange bias field exceeding 300 mT. Such sensitive manipulation is explained by the competition between the FM and AFM coupling at the interface of $CrI_3$ and $MnBi_2Te_4$, pointing to coverage-dependent interfacial magnetic interactions. Our work will facilitate the development of topological and antiferromagnetic devices.

Since the discovery of the quantum anomalous Hall (QAH) effect in magnetically doped topological insulators (TIs), researchers have devoted considerable effort to exploring new physics with the help of this novel state [1, 2]. To avoid random magnetic dopants, the intrinsic antiferromagnetic (AFM) TI [3-10] $MnBi_2Te_4$ was proposed. As a layered van der Waals (vdW) material, each $MnBi_2Te_4$ layer is a septuple-layer (SL), and two neighboring layers are weakly stacked by vdW forces (the lower inset of **Figure 1b**). Within each SL, the $Mn^{2+}$ moments are ferromagnetically aligned along the [0001] direction, whereas $Mn^{2+}$ moments belonging to adjacent SLs are antiferromagnetically coupled. In subsequent experimental works, the realization of zero-field QAH state [11] and several other intriguing topological states [12-16] in exfoliated $MnBi_2Te_4$ thin flakes confirmed this prediction.

However, it is suspected that the magnetic moments on the top surface of $MnBi_2Te_4$ deviates from the ideal situation [2, 17, 18], often resulting in the elimination of the magnetic topological gap and the absence of the QAH state. Improving the reproducibility and quality of the zero-field QAH state via surface/interface engineering of $MnBi_2Te_4$ is therefore a primary objective for the community. Indeed, (magnetic) TIs proximitized by magnetic insulators [19, 20] or superconductors [21, 22] typically exhibit intriguing physical phenomena. According to the theoretical prediction, proximity to a magnetic insulator such as $CrI_3$ could facilitate the formation of the QAH state in $MnBi_2Te_4$, which results from the large exchange interaction at the interface of $MnBi_2Te_4$ and $CrI_3$. [23] In addition, because both $MnBi_2Te_4$ and $CrI_3$ possess a sizable magnetic anisotropy with an out-of-plane easy axis [3, 24-26], the heterostructure device of $CrI_3/MnBi_2Te_4$ is a promising system and deserves investigation.

In this work, we fabricate FMI/AFM-TI vdW heterostructure devices of $CrI_3$/$MnBi_2Te_4$ by the dry transfer technique [27-29] and study their magnetotransport. Although the zero-field QAH effect is absent, the lateral shift of hysteresis loops, also known as the exchange bias effect, is successfully observed. Both negative and positive exchange bias effects are revealed by changing the interface configuration of $CrI_3$/$MnBi_2Te_4$ devices. Our work demonstrates the interesting magnetic coupling phenomenon at the interface of $CrI_3$/$MnBi_2Te_4$ and offers new insights for magnetic engineering of topological and antiferromagnetic devices.

Few-SLs $MnBi_2Te_4$ and nanoflakes of $CrI_3$ have been mechanically exfoliated from their bulk crystals. After the $MnBi_2Te_4$ field-effect transistor (FET) device is fabricated, the $CrI_3$/$MnBi_2Te_4$ heterostructure is formed by stacking the $CrI_3$ nanoflake on top of the $MnBi_2Te_4$ device. The thickness of $CrI_3$ nanoflakes is approximately 20 nm in this study, which is thin enough to be stacked above $MnBi_2Te_4$ to form an interface of satisfactory quality despite the presence of prepatterned electrodes. Meanwhile, the ~20 nm thick $CrI_3$ nanoflake can suppress thermal fluctuations and maintain a more stable magnetic anisotropy. More details about device fabrication are contained in the Experimental Methods section. **Figure 1a** shows the schematic of a $CrI_3$/$MnBi_2Te_4$ heterostructure device. As the $MnBi_2Te_4$ sample is completely covered by the $CrI_3$ nanoflakes, we call this configuration as the fully-covered configuration. Graphite and hexagonal boron nitride (h-BN) flakes are used to protect the $CrI_3$ and $MnBi_2Te_4$ samples from degradation [24].

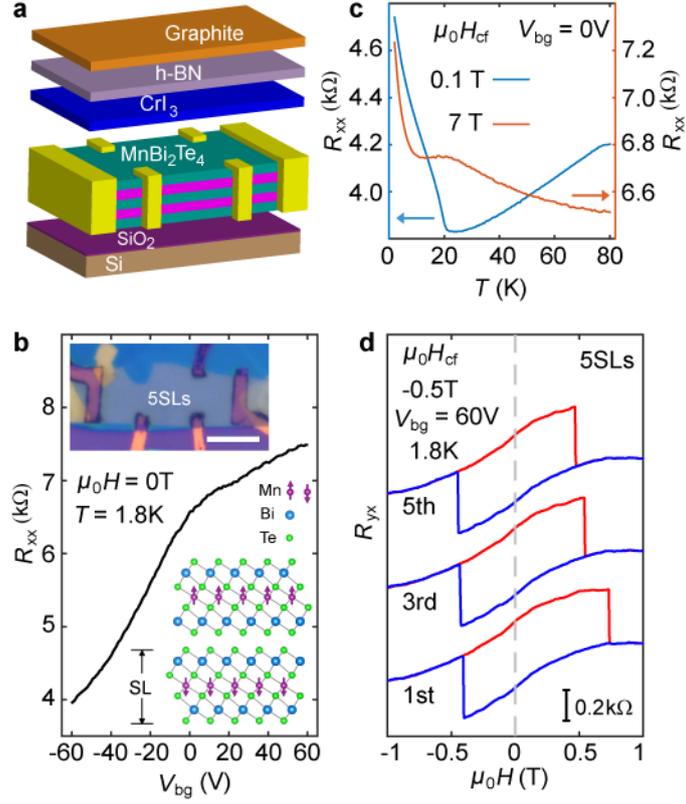

**Figure 1.** Exchange bias effect in $CrI_3$/$MnBi_2Te_4$ heterostructure devices. (a) The schematic of fully-covered device configuration. (b) The $V_{bg}$-dependent $R_{xx}$ of Device1 at zero magnetic field. The upper inset is the optical image of Device1. The lower inset is the atomic structure of $MnBi_2Te_4$. The scale bar in the top inset is 10 μm. (c) The temperature-dependent $R_{xx}$ curves of Device1 in field-cooling processes. (d) Exchange bias and training effects of Device1.

The upper inset of **Figure 1b** is an optical image of a $CrI_3$/(5-SLs)$MnBi_2Te_4$ device (labeled Device1) before the h-BN and graphite protective layers are stacked. **Figure 1b** depicts the back-gate voltage ($V_{bg}$)-dependent longitudinal resistance ($R_{xx}$) of Device1 at zero magnetic field, indicating that the carrier in this range is a hole type. To study the exchange bias effect, we employ the standard measurement methodology of field-cooling treatment [19, 20, 30]. Given that the Curie temperature ($T_C$) of bulk CrI$_3$ and the Néel temperature ($T_N$) of MnBi$_2$Te$_4$ crystals are 61 K and 24 K, respectively, we begin our measurement at 80 K and apply a polarizing magnetic field (cooling field, $\mu_0H_{cf}$) perpendicular to the sample plane. **Figure 1c** shows the temperature-dependent $R_{xx}$ during the field-cooling processes. As previously reported, the dip or hump at ~21 K in the $R_{xx}$-$T$ curves represents the Néel temperature of MnBi$_2$Te$_4$ layers [11, 31]. When the sample is cooled, we measure the magnetic field ($\mu_0H$)-dependent Hall resistance ($R_{yx}$), with the $\mu_0H$ perpendicular to the sample plane as well. The first (1st) curve in **Figure 1d** is the typical $R_{yx}$ loop measured after field-cooling with $\mu_0H_{cf}$ = -0.5 T of Device1, which exhibits a striking lateral shift along the horizontal axis toward the positive field. **Figure 1d** also shows the training behaviors, i.e., the exchange bias effect gradually disappears when we repeatedly sweep the hysteresis loop after field-cooling [32, 33]. Given the training effect, all $R_{yx}$ loops discussed in the rest of this paper are the first loop measured after the field-cooling treatment.

**Figure 2** shows detailed studies of the exchange bias effect in fully-covered CrI$_3$/MnBi$_2$Te$_4$ devices. As a typical FM/AFM bilayer heterostructure, CrI$_3$/MnBi$_2$Te$_4$ devices can serve as an ideal host for the exchange bias effect [32-35], which has benefited various commercial applications [32, 36]. From **Figure 1d** and **Figures 2a-c**, we can determine the negative exchange bias effect by definition,[19, 20] since the direction of the lateral shift of the $R_{yx}$ loop is opposite to the direction of $\mu_0H_{cf}$.

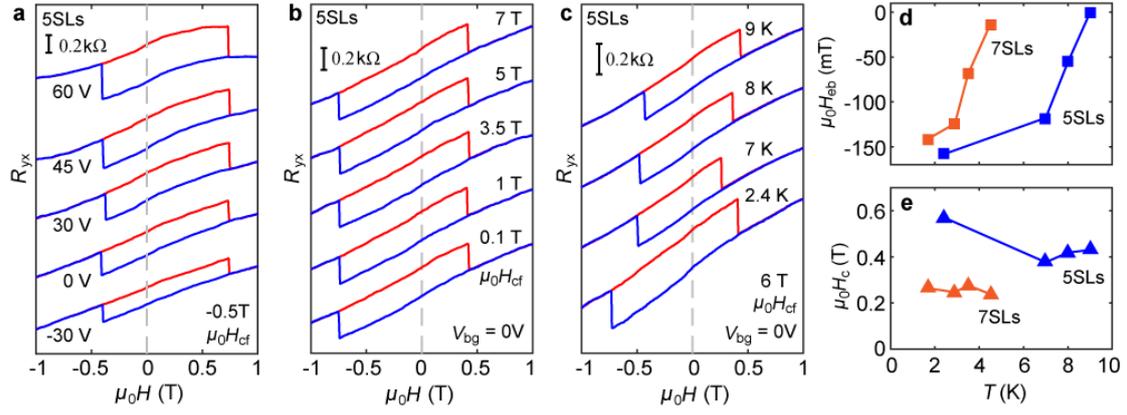

**Figure 2.** Back-gate voltage ($V_{bg}$), cooling field ($\mu_0 H_{cf}$) and temperature-dependent negative exchange bias effect in fully-covered $CrI_3$/$MnBi_2Te_4$ devices. (a, b) With a certain $\mu_0 H_{cf}$ ($V_{bg}$), the magnitude and sign of the exchange bias effect measured under different values of $V_{bg}$ ($\mu_0 H_{cf}$) are almost unchanged in Device1. The temperature is 1.8 K. (c) The exchange bias effect at different temperatures. With increasing temperature, the exchange bias disappears gradually. (d, e) Temperature-dependent $\mu_0 H_{eb}$ (d) and $\mu_0 H_c$ (e) of different devices (Device1 and Device2) with different $MnBi_2Te_4$ thicknesses.

To study whether the exchange bias field ($\mu_0 H_{eb}$ = (right coercive field + left coercive field)/2), which is a quantitative description of the lateral shift of the hysteresis loop [32, 35], is affected by the carrier density, we measure the exchange bias effect under different values of $V_{bg}$ (**Figure 2a**). Interestingly, in the $V_{bg}$ ranges that we exam, $\mu_0 H_{eb}$ shows only negligible fluctuations. **Figure 2b** shows the exchange bias effect of Device1 measured after field-cooling at various $\mu_0 H_{cf}$. As $\mu_0 H_{cf}$ is positive, the resulting $R_{yx}$ loop always shifts toward a negative field, and the magnitudes of the lateral shift are quite stable, regardless of the magnitudes of $\mu_0 H_{cf}$.

To explore the blocking temperature, i.e., the temperature above which the exchange bias effect disappears [35], we carry out the measurements by varying the temperature. **Figure 2c** shows the results of Device1. The temperature-dependent $\mu_0 H_{eb}$ of Device1 is extracted and plotted in **Figure 2d**. Its blocking temperature is 9 K, which is below the Néel temperature of MnBi$_2$Te$_4$ thin flakes (~21 K). The coercive field ($\mu_0 H_c$) is defined as (right coercive field − left coercive field)/2. **Figure 2e** shows the corresponding temperature dependence of $\mu_0 H_c$. Such temperature-dependent evolution of $\mu_0 H_{eb}$ and $\mu_0 H_c$ for another heterostructure device with a MnBi$_2$Te$_4$ thickness of 7 SLs (labeled Device2) is shown in **Figures 2d** and **2e** at the same time. See **Figure S2** for more details on Device2. Despite the obvious difference in $\mu_0 H_c$ between Device1 and Device2, the observed $\mu_0 H_{eb}$ at low temperature (~2 K) is at the same level (~ -150 mT).

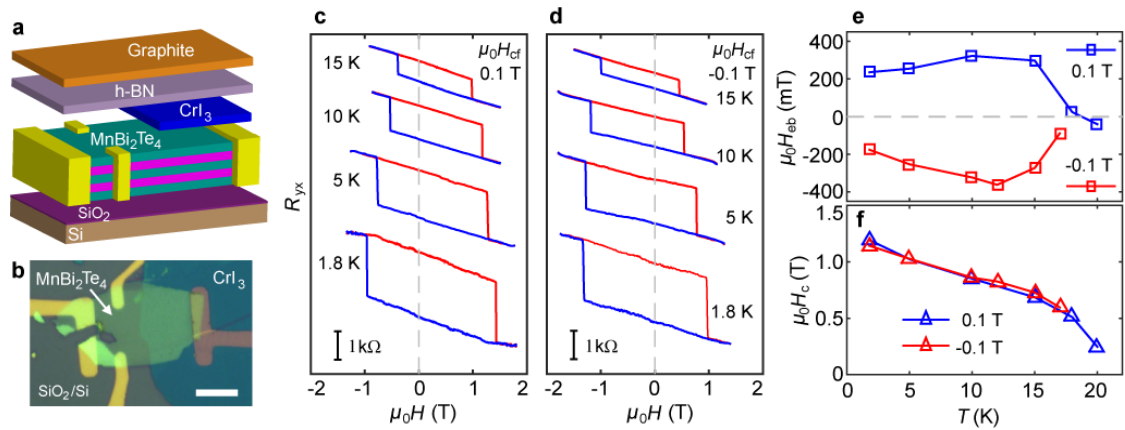

**Figure 3.** Positive exchange bias effect in half-covered CrI$_3$/MnBi$_2$Te$_4$ heterostructures. (a) The schematic of the device. (b) The optical image of Devie3 before the h-BN and graphite protective layers are stacked. The scale bar is 10 μm. (c, d) The temperature-dependent evolution of the positive exchange bias of Device3 field cooled at 0.1 T (c) and -0.1 T (d). The $V_{bg}$ is 0 V in these measurements. (e, f) The temperature-dependent

exchange bias field ($\mu_0H_{eb}$) and coercive field ($\mu_0H_c$) are summarized and plotted. $\mu_0H_{eb}$ and $\mu_0H_c$ of Device3 are enhanced compared to that of Device1 and Device2. The half-covered device has a higher blocking temperature.

When we change the interface configuration of CrI$_3$/MnBi$_2$Te$_4$ heterostructure devices by adjusting coverage approaches of CrI$_3$ nanoflake, as illustrated in **Figure 3a**, where only about half of MnBi$_2$Te$_4$ is covered by CrI$_3$, an interesting exchange bias effect appears. In particular, $R_{yx}$ is measured on the bare MnBi$_2$Te$_4$ region that is not covered by the CrI$_3$ layer. **Figure 3b** is an optical image of a fabricated device (labeled Device3), where the thickness of the MnBi$_2$Te$_4$ layer is 8 SLs and the nearly one-half coverage is obvious. Although the MnBi$_2$Te$_4$ thin flake is an even-layer sample in Device3, the hysteresis loop can also be observed as previously reported.[6, 11, 13, 37] **Figures 3c** and **3d** show the temperature-dependent exchange bias effect of Device3 field-cooled at 0.1 T and -0.1 T, respectively. The lateral shift direction of the $R_{yx}$ loops is consistent with the direction of $\mu_0H_{cf}$, indicating that it is a positive exchange bias effect,[19, 20] which is in contrast to the negative exchange bias effect observed in fully-covered CrI$_3$/MnBi$_2$Te$_4$ devices. The extracted $\mu_0H_{eb}$ versus temperature is plotted in **Figure 3e**, from which a blocking temperature of ~ 18 K can be determined, which is close to the Néel temperature of MnBi$_2$Te$_4$ thin flakes. Remarkably, $\mu_0H_{eb}$ could exceed 300 mT at a temperature ⩾10 K. We also note that, unlike fully-covered devices, the maximum $\mu_0H_{eb}$ does not occur at the lowest temperature for half-covered devices. The temperature-dependent $\mu_0H_c$ is summarized in **Figure 3f** as well. Obviously, both $\mu_0H_{eb}$ and $\mu_0H_c$ are greatly enhanced in the half-covered CrI$_3$/MnBi$_2$Te$_4$ devices. Similar observations can be seen in another half-covered device (labeled Device4, see **Figure S5**). It should be noted that the changes in MnBi$_2$Te$_4$ thickness can lead to the enhancement of $\mu_0H_c$ as well.[37]

Our observations are schematically summarized in **Figure 4**. The exchange bias effect originates from the unidirectional anisotropy induced by the exchange interaction between the FM and AFM layers at the heterostructure interface [20, 35]. Generally, in FM/AFM bilayer heterostructures, the negative (positive) exchange bias effect is a result of FM (AFM) interfacial coupling, i.e., magnetic moments belonging to different materials favor FM (AFM) alignment at the interface [19, 35, 38]. Based on this consideration, we conclude that the interfacial coupling in CrI$_3$/MnBi$_2$Te$_4$ heterostructures is coverage-dependent (or coverage-sensitive) and FM (AFM) type in fully-covered (half-covered) configuration **(Figure 4a-c)**. Although FM type interfacial coupling always demonstrates a negative exchange bias effect, AFM type interfacial coupling typically demonstrates a positive (negative) exchange bias effect when the $\mu_0H_{cf}$ is high (low).[19, 38] In the field-cooling process, the FM ordering of CrI$_3$ is formed as the temperature is lower than ~ 61 K. When the temperature is further lower than ~21 K, the surface spins of MnBi$_2$Te$_4$ will be ordered parallel to those of CrI$_3$ due to FM type coupling in fully-covered situation (**Figure 4a**). However, the surface spins of MnBi$_2$Te$_4$ can be ordered differently in half-covered situation. When $\mu_0H_{cf}$ is low, the exchange coupling energy between the FM and AFM layer predominates over the Zeeman energy induced by $\mu_0H_{cf}$ and vice versa.[19] Therefore, below ~21 K, the surface spins of MnBi$_2$Te$_4$ can be ordered antiparallel (parallel) to those of CrI$_3$ with a low (high) $\mu_0 H_{cf}$ (**Figure 4b, c**). But we cannot conclude from our experiments that the MnBi$_2$Te$_4$ surface spins can be perfectly aligned out of plane, so we use hollow arrows to represent them. Observations of a negative exchange bias effect in fully-covered CrI$_3$/MnBi$_2$Te$_4$ devices and a positive exchange bias effect in half-covered devices correspond to the situation of **Figures 4a** and **4c**, respectively. In the control experiment of another half-covered device (Device4), we also observed the situation of **Figure 4b** (for more details, see **Figure S6**).

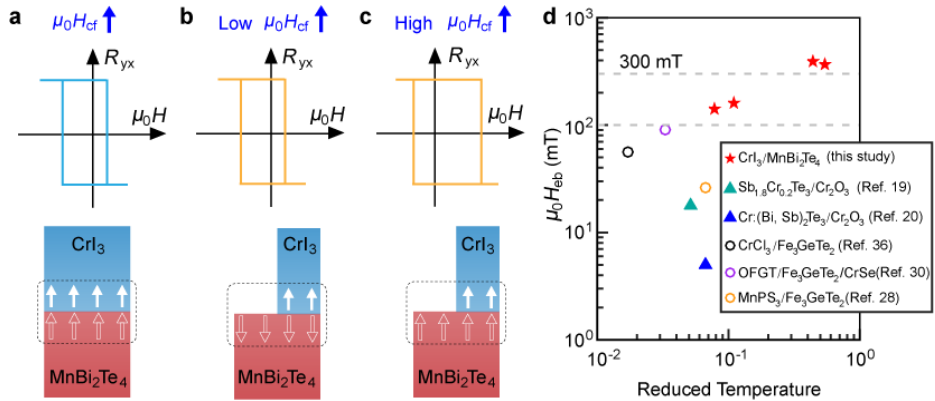

**Figure 4.** Coverage-dependent interfacial coupling and exchange bias effect with large $\mu_0 H_{eb}$ in CrI$_3$/MnBi$_2$Te$_4$ heterostructures. (a-c) Negative and positive exchange bias effects are observed in CrI$_3$/MnBi$_2$Te$_4$ heterostructures of different configurations, i.e., fully-covered (a) and half-covered (c) configurations. The simplistic illustrations of spin textures at the interface after field-cooling are shown as well. In the fully-covered (half-covered) device configuration, the spins of CrI$_3$ and MnBi$_2$Te$_4$ prefer FM (AFM) coupling at the interface. In a half-covered device configuration, as a result of AFM interfacial coupling strength competing against the Zeeman energy that originates from

$\mu_0H_{cf}$, the field-cooling processes with low $\mu_0H_{cf}$ (b) and high $\mu_0H_{cf}$ (c) can lead to different interfacial spin textures and exchange bias behavior. (d) The $\mu_0H_{eb}$ observed in this study and several other works are plotted. The four points of this study are the results of Device1-Device4 in the main text. The solid symbols are TIs related.

We note that the exchange bias effect reported in this study is revealed in the AFM layer ($MnBi_2Te_4$ here), which differs from usual circumstances [19, 20, 30, 35, 36]. The large $\mu_0H_{eb}$ observed in the $CrI_3/MnBi_2Te_4$ heterostructure is outstanding compared with several other vdW FM/AFM heterostructures studied in recent years. In **Figure 4d**, we provide a comparison of the $\mu_0H_{eb}$ observed in other works. As these works also measure $R_{yx}$, we define a reduced temperature as the measurement temperature divided by the Curie (or Néel) temperature of the material that contributes to the $R_{yx}$ signal. An alternative definition of the reduced temperature would be the measurement temperature divided by the blocking temperature, as suggested in **ref 35**. Since the blocking temperature is not explored in all prior studies, our definition is reasonable and informative for a qualitative comparison. Obviously, the large $\mu_0H_{eb}$ observed in $CrI_3/MnBi_2Te_4$ vdW heterostructures is striking.

These large $\mu_0H_{eb}$ and coverage-dependent interfacial interactions are explained by the competitive onset of FM and AFM coupling at the interface of $CrI_3$ and $MnBi_2Te_4$. Generally, interfacial coupling is either FM-type or AFM-type [19, 35]. However, the energy discrepancy between the FM and AFM interfacial coupling can sometimes be small, resulting in competitive switching between negative and positive exchange bias effects, as reported in protonic-gate-tuned oxidized Fe$_3$GeTe$_2$ nanoflakes [39]. For the CrI$_3$/MnBi$_2$Te$_4$ heterostructures studied here, FM coupling would be more favorable [23]. Therefore, in the fully-covered configuration, FM interfacial coupling and thus a negative exchange bias effect are anticipated. Although AFM coupling requires a higher energy in a fully-covered configuration, it is possible in a half-covered configuration because the competition between FM and AFM coupling is sensitive to interface details [40] and can be controlled. The larger $\mu_0H_{eb}$ in half-covered devices may indicate that the AFM coupling in this configuration is stronger than the FM coupling in the fully-covered configuration [32, 35]. The FM and AFM coupling competition picture helps us to understand our observations and suggests that minor changes at the interface of FM/AFM heterostructures can result in remarkable improvements in device performance.

One may also suspect that these observations are caused by stray fields [41] of the CrI$_3$ layer. In half-covered devices, the stray fields of the CrI$_3$ layer experienced by bare MnBi$_2$Te$_4$ region and CrI$_3$-covered MnBi$_2$Te$_4$ region have opposite directions. Such an explanation is ruled out by our control experiment in Device4, where the $R_{yx}$ of bare MnBi$_2$Te$_4$ region and CrI$_3$-covered MnBi$_2$Te$_4$ region are probed simultaneously. A positive exchange bias effect is observed in both regions. Moreover, the two regions have identical coercive fields and $\mu_0H_{eb}$ (see **Figure S5** for more details). This leads us to believe that the entire MnBi$_2$Te$_4$ layer is uniformly biased. In addition, given the smaller coercive field of CrI$_3$ compared to that of MnBi$_2$Te$_4$, it is unlikely that the stray fields of CrI$_3$ can bias the hysteresis loop of MnBi$_2$Te$_4$.

In summary, we demonstrate the interesting exchange bias effect in $CrI_3$/$MnBi_2Te_4$ vdW heterostructure devices. Depending on the heterostructure configuration, fully-covered and half-covered $CrI_3$/$MnBi_2Te_4$ devices exhibit negative and positive exchange bias effects, respectively. This coverage-dependent proximity approach for magnetic engineering would also work in other magnetic vdW heterostructures and deserves further investigation. The magneto-optical imaging techniques such as reflective magnetic circular dichroism (RMCD) imaging would provide a deeper comprehension of the observations here in the future. The large $\mu_0 H_{eb}$ in our experiment confirms the efficient interfacial coupling of $CrI_3$ and $MnBi_2Te_4$. Our work will facilitate the development of antiferromagnetic and TIs-based spintronics. [42]

**EXPERIMENTAL METHODS**

**Device Fabrication.** $MnBi_2Te_4$ bulk crystals were grown by the flux method as previously reported.[6] High-quality $CrI_3$ crystals were synthesized by a chemical vapor transport method using iodine ($I_2$) as a transport agent. Few-SLs $MnBi_2Te_4$ and thin flakes of $CrI_3$ (h-BN, graphite) were mechanically exfoliated onto 300-nm-thick $SiO_2$/Si substrate and polydimethylsiloxane (PDMS) substrate, respectively. The electrode pattern was defined by electron beam lithography, and then the Au film was evaporated to form electrical contact. After the lift-off process, the $MnBi_2Te_4$ FET device was sequentially covered by pre-exfoliated $CrI_3$, h-BN and graphite flakes. Most fabrication processes were carried out in a glove box with $H_2O$ and $O_2$ levels below 0.1 ppm.

**Transport Measurements.** The electrical transport measurements were conducted in a cryostat (1.6 K, 12 T), with the magnetic field applied perpendicular to the sample plane. The longitudinal resistance $R_{xx}$ and Hall resistance $R_{yx}$ were measured using standard lock-in technique with an a.c. excitation current of 50 nA at 13 Hz. The $R_{yx}$ data shown in **Figures 3c** and **3d** in the main text were anti-symmetrized to eliminate the mixing of $R_{yx}$ and $R_{xx}$ signals induced by the irregular sample geometry.[11,36] All other $R_{yx}$ data shown in the main text and **Supporting Information** are raw data, unless pointed out otherwise.

## ASSOCIATED CONTENT
### Supporting Information

The $R_{yx}$ curve of Device1 under high magnetic field; Chern insulator state and negative exchange bias effect of another fully-covered $CrI_3$/$MnBi_2Te_4$ heterostructure device (Device2); the $R_{yx}$ curve of Device3 under high magnetic field ; the raw data of Figures 3c and 3d; details of the positive exchange bias effect observed in Device4; the negative exchange bias effect in half-covered device field cooled at low $\mu_0H_{cf}$, i.e., the situation of Figure 4b; transport results in a $CrI_3$/$MnBi_2Te_4$ heterostructure device with a $CrI_3$-coverage between full-cover and half-cover.

## Author contributions

Z.Y. and B.C. contributed equally to this work. F.S., S.Z. and F.F. supervised the project. Z.Y. developed the heterostructure devices and performed the measurements. B.C. synthesized the crystals. Z.Y., S.Z., F.S., F.F., X.W., D.W. and H.Z. performed data analysis and discussion. Z.Y., S.Z. and F.S. wrote the paper with the help from all coauthors.

**Notes**

The authors declare no competing interests.

The data sets generated and/or analyzed during this study are available from the corresponding author upon reasonable request.

**ACKNOWLEDGEMENTS**

We gratefully acknowledge the financial support of the National Key R&D Program of China (No. 2022YFA1402404), the National Natural Science Foundation of China (Grant Nos. 92161201, T2221003, 12104221, 12025404, 12004174, 11904166, 11904165, 12274208, 61822403, and 11874203), the Natural Science Foundation of Jiangsu Province (Grant Nos. BK20200312 and BK20200310), and the Fundamental Research Funds for the Central Universities (Nos. 020414380192).